\documentclass[]{aastex63}

\graphicspath{{./}{figures/}}
\begin{document}

\title{A tentative emission line at z=5.8 from a 3\,mm-selected galaxy}


\correspondingauthor{Jorge A. Zavala}
\email{jzavala@utexas.edu}

\author[0000-0002-7051-1100]{Jorge A. Zavala}
\collaboration{1}{The University of Texas at Austin, 2515 Speedway Blvd, Austin, TX 78712, USA}


\begin{abstract}

I report a tentative ($\sim4\sigma$) emission line at $\nu=100.84\,$GHz from ``COS-3mm-1'', a 3\,mm-selected galaxy reported by \citet{Williams2019a} that is undetected at optical and near infrared wavelengths. The line was found in the ALMA Science Archive after re-processing ALMA band 3 observations targeting a different source. 
Assuming the line corresponds to the $\rm CO(6\to5)$ transition, this tentative detection implies a spectroscopic redshift of $z=5.857$, in  agreement with the  galaxy's redshift constraints from  multi-wavelength photometry. This would make this object the highest redshift 3\,mm-selected galaxy and one of the highest redshift dusty star-forming galaxies known to-date. Here, I report the characteristics of this tentative detection and the physical properties that can be inferred assuming the line is real. Finally, I advocate for follow-up observations to corroborate this identification and to confirm the high-redshift nature of this optically-dark dusty star-forming galaxy.

\end{abstract}


\keywords{ \hspace{0.2cm} High-redshift galaxies --- 
Redshift surveys --- Galaxy evolution  --- Molecular gas --- \hspace{0.4cm} Millimeter Astronomy --- Ultraluminous infrared galaxies\\}


\section{INTRODUCTION} 

The three millimeter window has recently been exploited to search for high-redshift dusty star-forming galaxies (DSFGs) through the detection of their dust continuum emission (e.g. \citealt{Zavala2018c,Zavala2021a,Gonzalez-Lopez2019a}). Thanks to the negative $K$-correction at this wavelength, 3\,mm-selected galaxies are expected to lie at $z>2$ (at the depth of the current surveys; \citealt{Casey2018b}).
The current searches have been motivated by the deep continuum maps obtained after collapsing spectroscopic observations, which are typically conducted in this waveband when targeting CO emission lines in high-redshift galaxies. ``COS-3mm-1', the source studied in this work  ($\alpha=\,$10:02:36.8, $\delta=\,$+02:08:40.6), was serendipitously discovered $\sim25''$ away from the phase center of an ALMA band 3 observation aimed at detecting CO in a galaxy at $z\sim1.5$ (\citealt{Williams2019a}).  

The galaxy was detected at $8\sigma$ with a 3\,mm flux density of $S_{\rm3mm}=150\rm\,\mu Jy$, and it is just marginally detected at the $\approx2-3\sigma$ level in {\it Spitzer}/IRAC, SCUBA-2 850$\,\mu$m, and VLA 3\,GHz  (after deblending the original maps with the ALMA positional prioir). The system drops out of deep optical imaging, including {\it HST}/CANDELS, Subaru, UltraVISTA, and it is also undetected in {\it Spitzer}/MIPS and in all the {\it Herschel} bands (\citealt{Williams2019a}). All these photometry datapoints were used by \citet{Williams2019a} to constrain the SED of the galaxy and its photometric redshift. 
The SED-fitting results suggest a high-redshift solution of $z_{\rm phot}=5.5^{+1.2}_{-1.1}$, with a 90\% probability of lying at $z>4.1$.\\

\section{ARCHIVAL DATA} 



With the aim of finding additional data to further constrain the redshift of this galaxy, I search for public ALMA observations around the position of ``COS-3mm-1'' using the ALMA Science Archive Query. Besides the original data in which the source was identified (ALMA project code: 2018.1.01739.S; PI: C. Williams), an additional program (2015.1.00861.S; PI: J. Silverman)
was publicly available in the archive (see \citealt{Silverman2018a} for details on the data).
These ALMA observations are centered $\sim40''$ away from the position of the source of interest. The primary beam response at the position of  ``COS-3mm-1" is around $0.29$. The data, which cover the frequencies $85.5-89.4\,$GHz and $97.7-101.5\,$GHz,  were re-analyzed following the standard ALMA pipeline scripts. Then, to search for emission lines, the spectral windows were imaged using natural weighting of the visibilities in order to maximize the signal-to-noise ratio (SNR).
The RMS reached over $100\rm\,km\,s^{-1}$ channel width was around 0.05\,mJy/beam.\\

\section{RESULTS} 

\begin{figure}[t]
\begin{center}
\vspace{0.5cm}
\includegraphics[scale=0.75]{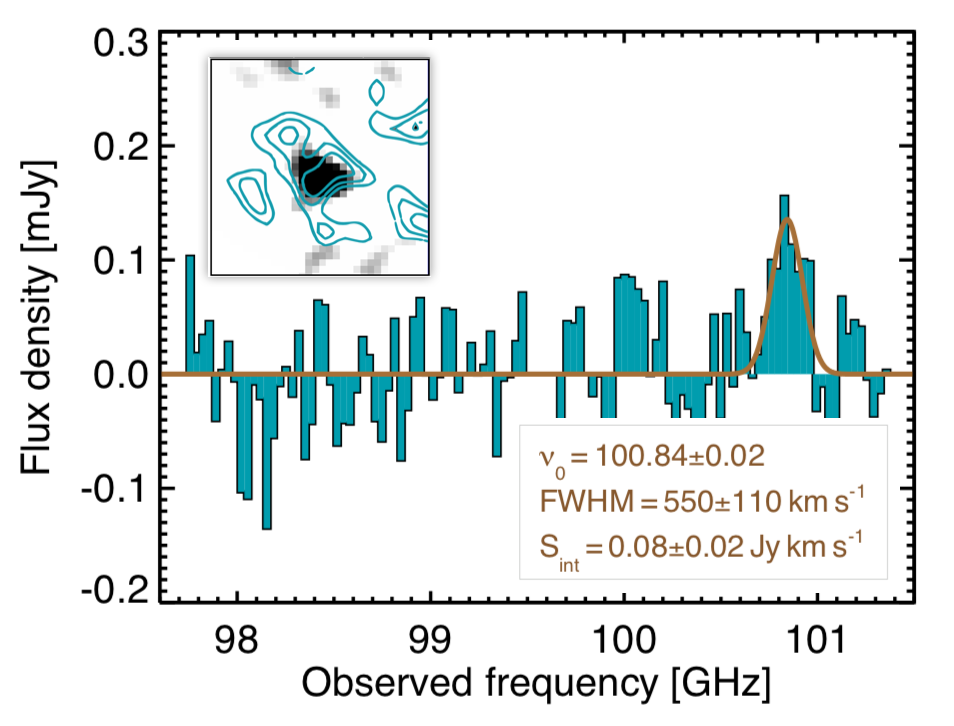}
\caption{Band 3 spectrum extracted at the position of ``COS-3mm-1''. The orange line represents the best-fit Gaussian function to the tentative emission line.  The center frequency of the line, the line-width, and the integrated line flux density are also indicated in the figure. The inset (black and white) plot shows the 3\,mm continuum detection from \citet{Williams2019a}  with  blue contours indicating the moment zero map of the line at 1, 2, and 3$\sigma$.
\label{fig1}}
\end{center}
\end{figure}

A tentative emission line was found at $\nu=110.84$\,GHz (Figure \ref{fig1}). The peak SNR is estimated to be around $3.0$, while the SNR of the integrated line is $\approx4.7$. The line is relatively well fit with a Gaussian function with line-width of $550\pm110\,\rm km\,s^{-1}$ and  integrated flux density of $0.08\pm0.02\,\rm Jy\,km\,s^{-1}$.  As revealed by the line moment-0 map shown in Figure \ref{fig1}, The spatial location of the line is coincident with the 3\,mm continuum detection from \citet{Williams2019a}, implying that, if real, this emission line arises from the 3\,mm-selected galaxy. 
Assuming the line is real and that it corresponds to a transition of carbon monoxide (the most common emission line in DSFGs), the most likely redshift solutions based on its photometric redshift constraints are $z=4.715$, $5.857$, and $6.999$, corresponding to the $\rm ^{12}CO(5\to4)$, $\rm ^{12}CO(6\to5)$, and $\rm ^{12}CO(7\to6)$ transitions, respectively. Here, I adopt $z=5.857\pm0.001$\footnote{Interestingly, ``MAMBO-9'', another DSFGs $\sim0.5$\,deg away from the source studied in this work (equivalent to around $10\rm\,Mpc$), lies at $z=5.850$ (\citealt{Casey2019a}). The confirmation of more galaxies with similar redshifts within this region of the sky might imply the existence of a large-scale galaxy proto-cluster structure. } as the redshift of the source since it is the closest solution to the maximum of the posterior redshift distribution found by \citealt{Williams2019a}.\\


The $^{12}\rm CO(6\to5)$ line luminosity of $L'_{\rm CO(6\to5)}=(2.55\pm 0.52)\times10^{9}\,\rm K\,km\,s^{-1}\,pc^2$ can be used to calculate the molecular gas mass of ``COS-3mm-1'' by, first, estimating the $\rm CO(1\to0)$ line luminosity and, then, assuming a CO-to-H$_2$ conversion factor. I assume the  line luminosity ratio reported by \citet{Bothwell2013a} ($L'_{\rm CO(6\to5)}/L'_{\rm CO(1\to0)}=0.21\pm0.04$) and a CO-to-H$_2$ conversion factor of $\alpha_{\rm CO}=4.6\,\rm M_\odot\,K\,km\,s^{-1}\,pc^2$. Based on these assumptions, I infer a gas mass of $M_{\rm gas}= (6\pm 2)\times 10^{10}\rm M_\odot$, which implies a depletion time scale ($t_{\rm dep}\equiv M_{\rm gas}/\rm SFR$) of $\sim200\rm\,Myr$ (adopting the star formation rate of $\sim300\,\rm M_\odot\,yr^{-1}$ derived by \citealt{Williams2019a}). This value is higher than the typical gas depletion timescales estimated for the bright population of high-redshift DSFGs, with values on the order of a few tens of Myrs (e.g. \citealt{Aravena2016a}). Hence,  ``COS-3mm-1'' is probably undergoing a long-lived star formation phase instead of the typical short-lived starburst episodes triggered by galaxy mergers (see discussion by \citealt{Jimenez2020a}).  This relatively long gas depletion timescale is in better agreement with those estimated for isolated star-forming disks at lower redshifts, and  might imply that the population of faint DSFGs (with $\rm SFR\lesssim500\rm\,M_\odot\,yr^{-1}$) forms stars through a smooth star formation mode over hundreds of Myrs. 

Nevertheless, the low significance of this detection prevents a robust confirmation of the redshift of this source and its physical properties. 
Follow-up observations are thus required not only to confirm this line, but also to identify a second line to unambiguously constrain its redshift. At $z=5.85$ the $^{12}\rm CO(5\to4)$ transition is redshifted to $\sim84.0$\,GHz, very close to the low frequency edge of the ALMA Band 3, making the detection unfeasible. The $^{12}\rm CO(5\to4)$ is better suited for NOEMA (NOrthern Extended Millimeter Array) or LMT (Large Millimeter Telescope) observations, although its detection might require exposure times of several hours. The $^{12}\rm CO(7\to6)$ transition is redshifted to $\sim117.6\,$GHz, a frequency covered only by the NOEMA interferometer. 
The [CII]-$158\mu$m transition will be observed at $\sim277.2\,$GHz, within the ALMA Band 7 coverage.  Given that the [CII] is one of the strongest FIR lines, its detection represents a promising way to confirm the redshift of this source. Other emission lines like the [OIII]-$88\mu$m or the [OI]-$63\mu$m, which have previously been  detected in high-redshift galaxies (\citealt{Hashimoto2018a,Rybak2020a}), also lie within the ALMA coverage, but their expected line luminosities are more uncertain and the lines are redshifted into the high-frequency ALMA bands which require very good observing conditions. 
Alternatively, the redshift of this source can be confirmed with JWST observations, through the  detection of rest-frame ultraviolet and optical emission lines, such as  Ly$\alpha$, [OII]$3727\AA$, H$\beta$, [OIII]$4959,5007\AA$, and/or H$\alpha$.\\

\section{CONCLUSIONS} 

The tentative emission line reported in this work suggests that, if real, ``COS-3mm-1'' is the highest redshift galaxy selected at 3\,mm and one of the highest DSFGs known to-date. The only three DSFGs identified via (sub-)millimeter observations with higher spectroscopic redshifts are all gravitationally lensed and, with the exception of G09-83808 (\citealt{Zavala2018a}), they are rare extreme galaxies with SFRs exceeding $1000\,\rm M_\odot\,yr^{-1}$ (\citealt{Riechers2013a,Strandet2017a}). Therefore, this source could, potentially, serve as an important benchmark for $z>5$ DSFG studies for the whole extragalactic community, particularly, if it is confirmed to be at the same redshift as `MAMBO-9' (\citealt{Casey2019a}), which would suggest the existence of a galaxy proto-cluster at $z=5.85$ in this field. The confirmation of its redshift and its physical properties is thus imperative.



\acknowledgments

Jorge A.  Zavala thanks Christina Williams, Caitlin Casey, and Sinclaire Manning for their feedback on this manuscript.

\section*{DATA AVAILABILITY}
This paper makes use of the following ALMA data:
ADS/JAO.ALMA\#2015.1.00861.S and ADS/JAO.ALMA
\#2018.1.01739.S,
archived at \url{https://almascience.nrao.edu/alma-data/archive}.

\newpage 

\bibliography{sample63}{}

\begin{thebibliography}{}
\expandafter\ifx\csname natexlab\endcsname\relax\def\natexlab#1{#1}\fi
\providecommand{\url}[1]{\href{#1}{#1}}
\providecommand{\dodoi}[1]{doi:~\href{http://doi.org/#1}{\nolinkurl{#1}}}
\providecommand{\doeprint}[1]{\href{http://ascl.net/#1}{\nolinkurl{http://ascl.net/#1}}}
\providecommand{\doarXiv}[1]{\href{https://arxiv.org/abs/#1}{\nolinkurl{https://arxiv.org/abs/#1}}}

\bibitem[{{Aravena} {et~al.}(2016){Aravena}, {Spilker}, {Bethermin},
  {Bothwell}, {Chapman}, {de Breuck}, {Furstenau}, {G{\'o}nzalez-L{\'o}pez},
  {Greve}, {Litke}, {Ma}, {Malkan}, {Marrone}, {Murphy}, {Stark}, {Strandet},
  {Vieira}, {Weiss}, {Welikala}, {Wong}, \& {Collier}}]{Aravena2016a}
{Aravena}, M., {Spilker}, J., {Bethermin}, M., {Bothwell}, M., {Chapman}, S.,
  {et~al.} 2016, \mnras, 457, 4406, \dodoi{10.1093/mnras/stw275}

\bibitem[{{Bothwell} {et~al.}(2013){Bothwell}, {Smail}, {Chapman}, {Genzel},
  {Ivison}, {Tacconi}, {Alaghband-Zadeh}, {Bertoldi}, {Blain}, {Casey}, {Cox},
  {Greve}, {Lutz}, {Neri}, {Omont}, \& {Swinbank}}]{Bothwell2013a}
{Bothwell}, M., {Smail}, I., {Chapman}, S., {Genzel}, R., {Ivison}, R.~J.,
  {et~al.} 2013, \mnras, 429, 3047, \dodoi{10.1093/mnras/sts562}

\bibitem[{{Casey} {et~al.}(2018){Casey}, {Hodge}, {Zavala}, {Spilker}, {da
  Cunha}, {Staguhn}, {Finkelstein}, \& {Drew}}]{Casey2018b}
{Casey}, C., {Hodge}, J., {Zavala}, J., {Spilker}, J., {da Cunha}, E., {et~al.}
  2018, \apj, 862, 78, \dodoi{10.3847/1538-4357/aacd11}

\bibitem[{{Casey} {et~al.}(2019){Casey}, {Zavala}, {Aravena}, {B{\'e}thermin},
  {Caputi}, {Champagne}, {Clements}, {da Cunha}, {Drew}, {Finkelstein},
  {Hayward}, {Kartaltepe}, {Knudsen}, {Koekemoer}, {Magdis}, {Man}, {Manning},
  {Scoville}, {Sheth}, {Spilker}, {Staguhn}, {Talia}, {Taniguchi}, {Toft},
  {Treister}, \& {Yun}}]{Casey2019a}
{Casey}, C., {Zavala}, J., {Aravena}, M., {B{\'e}thermin}, M., {Caputi}, K.~I.,
  {et~al.} 2019, \apj, 887, 55, \dodoi{10.3847/1538-4357/ab52ff}

\bibitem[{{Gonz{\'a}lez-L{\'o}pez} {et~al.}(2019){Gonz{\'a}lez-L{\'o}pez},
  {Decarli}, {Pavesi}, {Walter}, {Aravena}, {Carilli}, {Boogaard}, {Popping},
  {Weiss}, {Assef}, {Bauer}, {Bertoldi}, {Bouwens}, {Contini}, {Cortes}, {Cox},
  {da Cunha}, {Daddi}, {D{\'\i}az-Santos}, {Inami}, {Hodge}, {Ivison}, {Le
  F{\`e}vre}, {Magnelli}, {Oesch}, {Riechers}, {Rix}, {Smail}, {Swinbank},
  {Somerville}, {Uzgil}, \& {van der Werf}}]{Gonzalez-Lopez2019a}
{Gonz{\'a}lez-L{\'o}pez}, J., {Decarli}, R., {Pavesi}, R., {Walter}, F.,
  {Aravena}, M., {et~al.} 2019, \apj, 882, 139,
  \dodoi{10.3847/1538-4357/ab3105}

\bibitem[{{Hashimoto} {et~al.}(2018){Hashimoto}, {Laporte}, {Mawatari},
  {Ellis}, {Inoue}, {Zackrisson}, {Roberts-Borsani}, {Zheng}, {Tamura},
  {Bauer}, {Fletcher}, {Harikane}, {Hatsukade}, {Hayatsu}, {Matsuda}, {Matsuo},
  {Okamoto}, {Ouchi}, {Pell{\'o}}, {Rydberg}, {Shimizu}, {Taniguchi},
  {Umehata}, \& {Yoshida}}]{Hashimoto2018a}
{Hashimoto}, T., {Laporte}, N., {Mawatari}, K., {Ellis}, R.~S., {Inoue}, A.~K.,
  {et~al.} 2018, \nat, 557, 392, \dodoi{10.1038/s41586-018-0117-z}

\bibitem[{{Jim{\'e}nez-Andrade} {et~al.}(2020){Jim{\'e}nez-Andrade}, {Zavala},
  {Magnelli}, {Casey}, {Liu}, {Romano-D{\'\i}az}, {Schinnerer}, {Harrington},
  {Aretxaga}, {Karim}, {Staguhn}, {Burnham}, {Monta{\~n}a},
  {Smol{\v{c}}i{\'c}}, {Yun}, {Bertoldi}, \& {Hughes}}]{Jimenez2020a}
{Jim{\'e}nez-Andrade}, E., {Zavala}, J., {Magnelli}, B., {Casey}, C.~M., {Liu},
  D., {et~al.} 2020, \apj, 890, 171, \dodoi{10.3847/1538-4357/ab6dec}

\bibitem[{{Riechers} {et~al.}(2013){Riechers}, {Bradford}, {Clements},
  {Dowell}, {P{\'e}rez-Fournon}, {Ivison}, {Bridge}, {Conley}, {Fu}, {Vieira},
  {Wardlow}, {Calanog}, {Cooray}, {Hurley}, {Neri}, {Kamenetzky}, {Aguirre},
  {Altieri}, {Arumugam}, {Benford}, {B{\'e}thermin}, {Bock}, {Burgarella},
  {Cabrera-Lavers}, {Chapman}, {Cox}, {Dunlop}, {Earle}, {Farrah}, {Ferrero},
  {Franceschini}, {Gavazzi}, {Glenn}, {Solares}, {Gurwell}, {Halpern},
  {Hatziminaoglou}, {Hyde}, {Ibar}, {Kov{\'a}cs}, {Krips}, {Lupu}, {Maloney},
  {Martinez-Navajas}, {Matsuhara}, {Murphy}, {Naylor}, {Nguyen}, {Oliver},
  {Omont}, {Page}, {Petitpas}, {Rangwala}, {Roseboom}, {Scott}, {Smith},
  {Staguhn}, {Streblyanska}, {Thomson}, {Valtchanov}, {Viero}, {Wang},
  {Zemcov}, \& {Zmuidzinas}}]{Riechers2013a}
{Riechers}, D., {Bradford}, C., {Clements}, D., {Dowell}, C.,
  {P{\'e}rez-Fournon}, I., {et~al.} 2013, \nat, 496, 329,
  \dodoi{10.1038/nature12050}

\bibitem[{{Rybak} {et~al.}(2020){Rybak}, {Zavala}, {Hodge}, {Casey}, \&
  {Werf}}]{Rybak2020a}
{Rybak}, M., {Zavala}, J.~A., {Hodge}, J.~A., {Casey}, C.~M., \& {Werf}, P.
  v.~d. 2020, \apjl, 889, L11, \dodoi{10.3847/2041-8213/ab63de}

\bibitem[{{Silverman} {et~al.}(2018){Silverman}, {Rujopakarn}, {Daddi},
  {Renzini}, {Rodighiero}, {Liu}, {Puglisi}, {Sargent}, {Mancini},
  {Kartaltepe}, {Kashino}, {Koekemoer}, {Arimoto}, {B{\'e}thermin}, {Jin},
  {Magdis}, {Nagao}, {Onodera}, {Sanders}, \& {Valentino}}]{Silverman2018a}
{Silverman}, J., {Rujopakarn}, W., {Daddi}, E., {Renzini}, A., {Rodighiero},
  G., {et~al.} 2018, \apj, 867, 92, \dodoi{10.3847/1538-4357/aae25e}

\bibitem[{{Strandet} {et~al.}(2017){Strandet}, {Weiss}, {De Breuck}, {Marrone},
  {Vieira}, {Aravena}, {Ashby}, {B{\'e}thermin}, {Bothwell}, {Bradford},
  {Carlstrom}, {Chapman}, {Cunningham}, {Chen}, {Fassnacht}, {Gonzalez},
  {Greve}, {Gullberg}, {Hayward}, {Hezaveh}, {Litke}, {Ma}, {Malkan}, {Menten},
  {Miller}, {Murphy}, {Narayanan}, {Phadke}, {Rotermund}, {Spilker}, \&
  {Sreevani}}]{Strandet2017a}
{Strandet}, M., {Weiss}, A., {De Breuck}, C., {Marrone}, D., {Vieira}, J.~D.,
  {et~al.} 2017, \apjl, 842, L15, \dodoi{10.3847/2041-8213/aa74b0}

\bibitem[{{Williams} {et~al.}(2019){Williams}, {Labbe}, {Spilker}, {Stefanon},
  {Leja}, {Whitaker}, {Bezanson}, {Narayanan}, {Oesch}, \&
  {Weiner}}]{Williams2019a}
{Williams}, C.~C., {Labbe}, I., {Spilker}, J., {Stefanon}, M., {Leja}, J.,
  {et~al.} 2019, \apj, 884, 154, \dodoi{10.3847/1538-4357/ab44aa}

\bibitem[{{Zavala} {et~al.}(2018{\natexlab{a}}){Zavala}, {Casey}, {da Cunha},
  {Spilker}, {Staguhn}, {Hodge}, \& {Drew}}]{Zavala2018c}
{Zavala}, J., {Casey}, C., {da Cunha}, E., {Spilker}, J., {Staguhn}, J.,
  {et~al.} 2018{\natexlab{a}}, \apj, 869, 71, \dodoi{10.3847/1538-4357/aaecd2}

\bibitem[{{Zavala} {et~al.}(2018{\natexlab{b}}){Zavala}, {Monta{\~n}a},
  {Hughes}, {Yun}, {Ivison}, {Valiante}, {Wilner}, {Spilker}, {Aretxaga},
  {Eales}, {Avila-Reese}, {Ch{\'a}vez}, {Cooray}, {Dannerbauer}, {Dunlop},
  {Dunne}, {G{\'o}mez-Ruiz}, {Micha{\l}owski}, {Narayanan}, {Nayyeri}, {Oteo},
  {Rosa Gonz{\'a}lez}, {S{\'a}nchez-Arg{\"u}elles}, {Schloerb}, {Serjeant},
  {Smith}, {Terlevich}, {Vega}, {Villalba}, {van der Werf}, {Wilson}, \&
  {Zeballos}}]{Zavala2018a}
{Zavala}, J., {Monta{\~n}a}, A., {Hughes}, D., {Yun}, M., {Ivison}, R.,
  {et~al.} 2018{\natexlab{b}}, Nature Astronomy, 2, 56,
  \dodoi{10.1038/s41550-017-0297-8}

\bibitem[{{Zavala} {et~al.}(2021){Zavala}, {Casey}, {Manning}, {Aravena},
  {Bethermin}, {Caputi}, {Clements}, {da Cunha}, {Drew}, {Finkelstein},
  {Fujimoto}, {Hayward}, {Hodge}, {Kartaltepe}, {Knudsen}, {Koekemoer}, {Long},
  {Magdis}, {Man}, {Popping}, {Sanders}, {Scoville}, {Sheth}, {Staguhn},
  {Toft}, {Treister}, {Vieira}, \& {Yun}}]{Zavala2021a}
{Zavala}, J., {Casey}, C., {Manning}, S., {Aravena}, M., {Bethermin}, M.,
  {et~al.} 2021, arXiv e-prints, arXiv:2101.04734.
\newblock \doarXiv{2101.04734}

\end{thebibliography}
\bibliographystyle{aasjournal}






\end{document}